\documentclass[aps,prb,twocolumn,preprintnumbers,amsmath,amssymb,superscriptaddress, showpacs]{revtex4-1}

\usepackage{graphicx}% Include figure files
\usepackage{dcolumn}% Align table columns on decimal point
\usepackage{bm}% bold math
\usepackage{color}
\usepackage{textcase, natbib, url, ulem}
\usepackage[colorlinks=true, linkcolor=blue, urlcolor=blue, citecolor=blue]{hyperref}
\usepackage{amsmath,amssymb}
\usepackage{mathtools}
\definecolor{darkblue}{rgb}{0,0,0.5}
\definecolor{darkgreen}{rgb}{0,0.5,0}
\definecolor{darkred}{rgb}{.7,0,0}
\definecolor{purple}{rgb}{0.5,0,0.6}
\definecolor{orange}{rgb}{1,0.5,0}
\definecolor{grey}{rgb}{.6,.6,.6}
\definecolor{lightpink}{rgb}{1,0.7,0.75}
\definecolor{pink}{rgb}{1,0.4,0.58}
\definecolor{deeppink}{rgb}{1,0.08,0.58}

\newcommand{\cb}[1]{{\color{darkblue}{#1}}}

\begin{document}

\preprint{draft}
\title{Low-temperature behavior of transmission phase shift across a Kondo correlated quantum dot}% Force line breaks with \\

\author{S. Takada}
\altaffiliation[Present address:]{CNRS, Institut N\'{e}el, F-38042 Grenoble, France}
\affiliation{Department of Applied Physics, University of Tokyo, Bunkyo-ku, Tokyo, 113-8656, Japan}
\author{M. Yamamoto}
\affiliation{Department of Applied Physics, University of Tokyo, Bunkyo-ku, Tokyo, 113-8656, Japan}
\affiliation{PRESTO, JST, Kawaguchi-shi, Saitama 331-0012, Japan}
\author{C. B\"{a}uerle}
\affiliation{Universit\'{e}. Grenoble Alpes, Institut N\'{e}el, F-38042 Grenoble, France}
\affiliation{CNRS, Institut N\'{e}el, F-38042 Grenoble, France}
\author{A. Alex}
\affiliation{Physics Department, Arnold Sommerfeld Center for Theoretical Physics, and Center for NanoScience, Ludwig-Maximilians-Universit\"{a}t, Theresienstra\ss e 37, D-80333 M\"{u}nchen, Germany}
\author{J. von Delft}
\affiliation{Physics Department, Arnold Sommerfeld Center for Theoretical Physics, and Center for NanoScience, Ludwig-Maximilians-Universit\"{a}t, Theresienstra\ss e 37, D-80333 M\"{u}nchen, Germany}
\author{A. Ludwig}
\affiliation{Lehrstuhl f\"{u}r Angewandte Festk\"{o}rperphysik, Ruhr-Universit\"{a}t Bochum, Universit\"{a}tsstra\ss e 150, D-44780 Bochum, Germany}
\author{A. D. Wieck}
\affiliation{Lehrstuhl f\"{u}r Angewandte Festk\"{o}rperphysik, Ruhr-Universit\"{a}t Bochum, Universit\"{a}tsstra\ss e 150, D-44780 Bochum, Germany}
\author{S. Tarucha}
\affiliation{Department of Applied Physics, University of Tokyo, Bunkyo-ku, Tokyo, 113-8656, Japan}
\affiliation{RIKEN Center for Emergent  Matter Science (CEMS), 2-1 Hirosawa, Wako-shi, Saitama 31-0198, Japan}

\date{\today}% It is always \today, today,

\begin{abstract}
  We study the transmission phase shift across a Kondo correlated quantum dot in a GaAs heterostructure at temperatures below the Kondo temperature ($T < T_{\rm K}$), where the phase shift is expected to show a plateau at $\pi/2$ for an ideal Kondo singlet ground state.
  Our device is tuned such that the ratio $\Gamma/U$ of level width $\Gamma$ to charging energy $U$ is quite large ($\lesssim 0.5$ rather than $\ll 1$).
  This situation is commonly used in GaAs quantum dots to ensure Kondo temperatures large enough ($\simeq 100$ mK here) to be experimentally accessible; however it also implies that charge fluctuations are more pronounced than typically assumed in theoretical studies focusing on the regime $\Gamma/U \ll 1$ needed to ensure a well-defined local moment.
  Our measured phase evolves monotonically by $\pi$ across the two Coulomb peaks, but without being locked at $\pi/2$ in the Kondo valley for $T \ll T_{\rm K}$, due to a significant influence of large $\Gamma/U$.
  Only when $\Gamma/U$ is reduced sufficiently does the phase start to be locked around $\pi/2$ and develops into a plateau at $\pi/2$.
  Our observations are consistent with numerical renormalization group calculations, and can be understood as a direct consequence of the Friedel sum rule that relates the transmission phase shift to the local occupancy of
  the dot, and thermal average of a transmission coefficient through a resonance level near the Fermi energy.
\end{abstract}

\maketitle

%\tableofcontents

%%% Introduction %%%%%%%%%%%%%%
%\section{\label{sec:intro}Introduction}%%%%%%
%%%%%%%%%%%%%%%%%%%%%%%%
The Kondo effect \cite{JKondo1964} that is the archetype of many-body correlations was first observed in metals containing small inclusions of magnetic impurities and later in semiconductor quantum dots (QDs) \cite{GoldhaberGordon1998, Cronenwett1998} where a single electron spin can be artificially created.
The flexibility of various tuning parameters of a QD makes it an extremely favorable system for studying the Kondo effect, and several new properties such as the unitary limit of conductance \cite{vanderwiel2000} and non-equilibrium Kondo density of states (DOS) \cite{Franceschi2002, Leturcq2005} have been observed through conductance measurements.
Furthermore by embedding a QD into quantum interferometers it also becomes possible to investigate the transmission phase shift of electrons scattered through a Kondo correlated QD \cite{Ji2000, Ji2002, Zaffalon2008, Sato2005, Katsumoto2006, Takada2014}.
An electron scattered through a Kondo correlated QD is expected to acquire a $\pi/2$-phase shift without spin flip scattering \cite{Gerland2000} at zero temperature, which is a central ingredient of Nozi\`{e}res' Fermi-liquid theory \cite{Nozieres:xy}.
It is also a direct indication of a many-body singlet ground state, often referred to as the Kondo cloud.

Very recently, the $\pi/2$-phase shift in the Kondo regime has been unambiguously observed by measuring the phase through a QD embedded in a \textit{true} two-path interferometer \cite{Takada2014}. 
The difficulty in measuring the phase shift through a QD originates from the difficulty in realizing a \textit{pure} two-path interferometer in a mesoscopic system \cite{Yamamoto:2012fk, Tobias2014, Aharony2014, TakadaAPL2015} due to boundary conditions for linear transport \cite{Buttiker1988, Yeyati1995}.
It has been shown by some of us that the anti-phase oscillation in our interferometer works as a good criterion for reliable phase measurement while the smooth phase evolution as a function of gate voltages, which is the criterion used in previous experiments, does not \cite{TakadaAPL2015}.
%---------------
According to theory, it is also expected that the $\pi/2$-phase shift is washed out due to the influence of neighboring levels \cite{Gerland2000}.
Such a situation may indeed have been realized in earlier experiments for $T \ll T_{\rm K}$, but reliable phase measurements in this regime have stayed elusive due to technical difficulties.
It is therefore of interest to study this regime in order to obtain a comprehensive understanding of electron transport across a single Kondo impurity coupled to reservoirs at finite temperatures.
%--------------

Here we study the phase behavior of a Kondo correlated QD formed in a GaAs heterostructure at $T \lesssim T_{\rm K}$ by varying $\Gamma/U$.
We show that the level broadening has a strong influence on the phase shift evolution in the low temperature Fermi-liquid regime.
We find that the phase smoothly and monotonically shifts by $\pi$ across two Coulomb peaks (CPs) without being locked at $\pi/2$ at the center of the Kondo valley when there is a significant overlap of the levels.
When decreasing the level broadening $\Gamma/U$, the phase shift starts to be locked around $\pi/2$ and then eventually develops into a pronounced plateau.
We will argue below that this behavior, reproduced by numerical renormalization group (NRG) calculations, can be understood in terms of the Friedel sum rule, which relates the transmission phase shift to the dot occupancy, and a thermal average of the transmission phase.

%%%% Measurement setup %%%%%%%%
%
%%%%%%%%% Figure 1 %%%%%%%%%%%%%%%%%%%%
\begin{figure}[htbp]
	\includegraphics[width=0.48\textwidth]{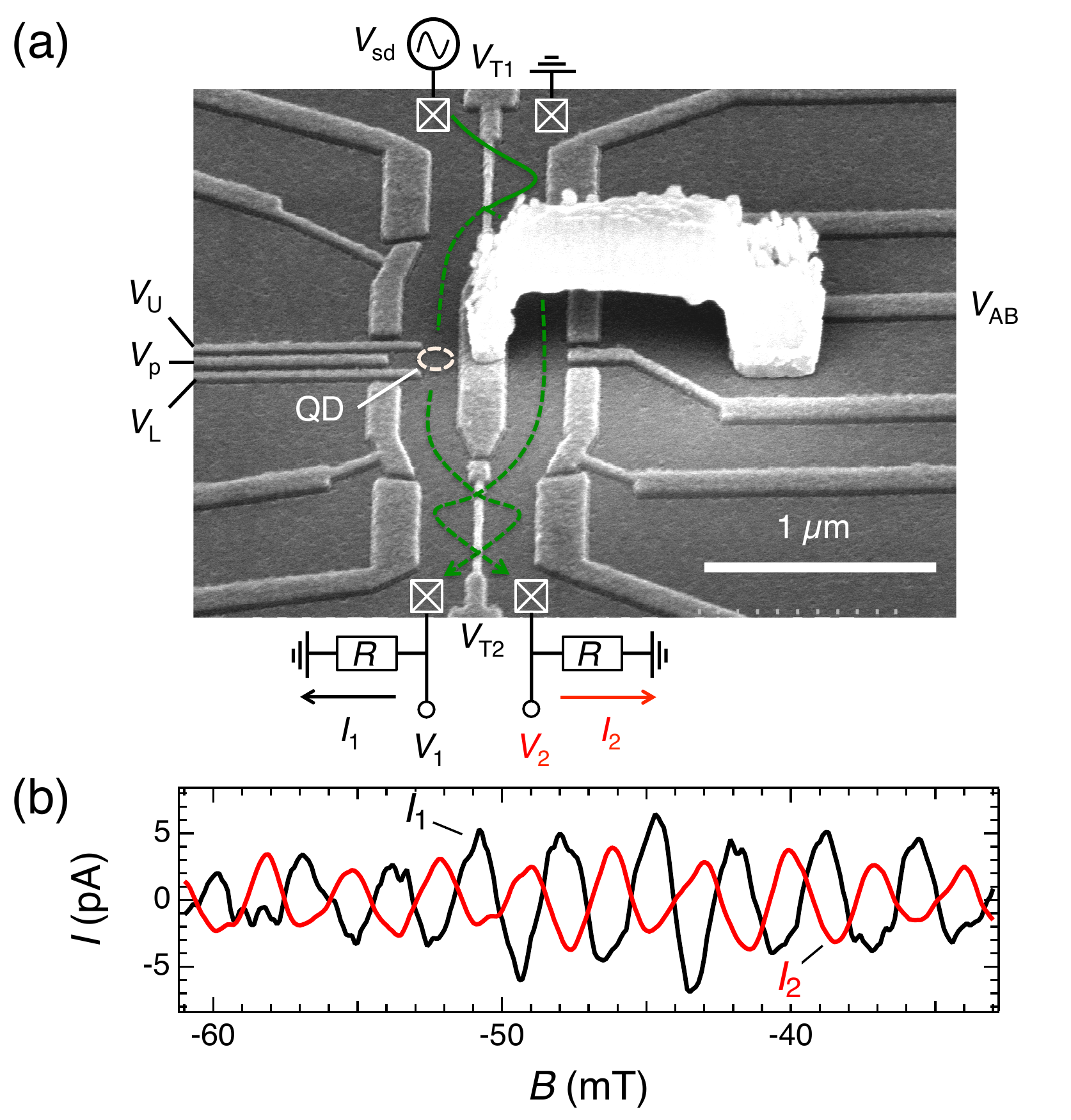}
	\caption{\label{fig:device}(a) A scanning electron micrograph of the device including the measurement setup. The dotted green lines mimic the electron trajectories. The currents flowing through the lower contacts are measured through voltage measurements [$I_{\rm 1(2)} = V_{\rm 1(2)}/R$, $R=10\ {\rm k\Omega}$]. (b) Typical quantum interference as a function of the magnetic field applied perpendicularly to the surface. Non-oscillating background current is subtracted from the raw data and only the oscillating component is shown. The visibilities of the oscillations are $4 \% \sim 8$ \%.}
\end{figure}
\begin{figure}[htbp]
	\includegraphics[width=0.48\textwidth]{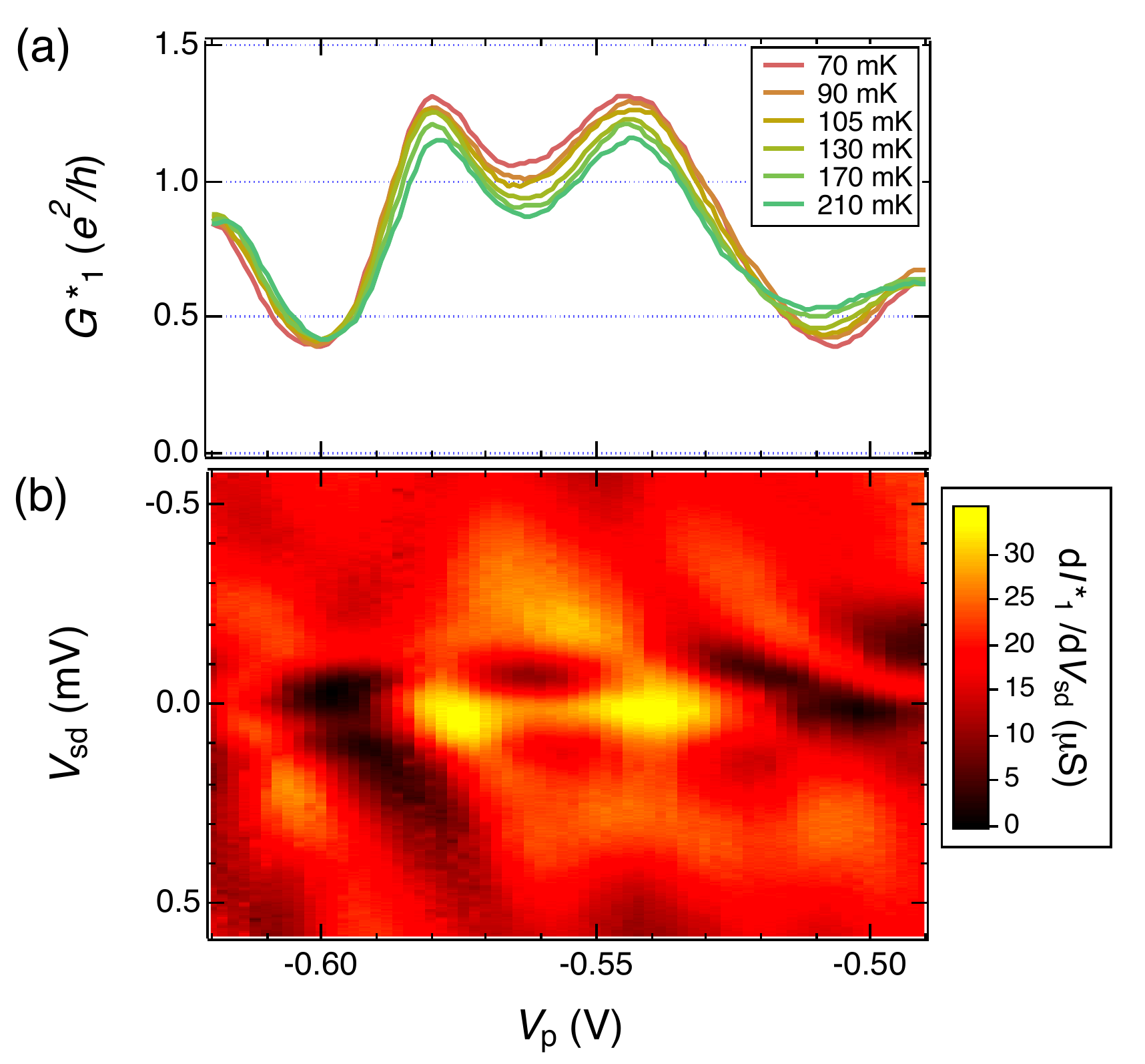}
	\caption{\label{fig:Kondo}(a) Temperature dependence of the Coulomb peaks. The conductance at different temperatures is plotted with different colors. (b) Coulomb diamond for the Coulomb peaks shown in (a) at $T = 70 \ {\rm mK}$.}
\end{figure}

%%%%%%%%%%%%%%%%%%%%%%%%%%%%%%%%%%%%
The device is fabricated from a GaAs/AlGaAs heterostructure that hosts a two-dimensional electron gas (2DEG) with $n=3.21 \times 10^{11}~{\rm cm^{-2}}$, $\mu=8.6 \times 10^{5} ~{\rm cm^{2}/Vs}$ located at 100 nm below the surface with a modulation doping and a 45 nm spacer between doping and 2DEG.
The sample structure [Fig.\,\ref{fig:device}(a)] is defined using a metallic Schottky gate technique.
The AB ring is defined by the gate voltage $V_{\rm AB}$ applied through a metallic air bridge onto the center island.
It is connected to the leads through a tunnel-coupled wire on both ends, where the tunnel-coupling energy is controlled by $V_{\rm T1}$ and $V_{\rm T2}$.
The metallic air bridge allows us to control the two tunnel-coupled wires independent on the AB ring, which improves the tunability of the interferometer compared to the previous setup \cite{Takada2014}.
A QD is formed in the left path of the AB ring and $V_{\rm p}$ is used to control the single-particle level of the QD.
A current is injected from the upper left contact by applying an ac bias ($V_{\rm sd} = 3 - 20~{\rm \mu V}$, $23.3$ Hz) and recovered in the two contacts at the bottom.
The currents ($I_{\rm 1}$, $I_{\rm 2}$) are measured as voltages ($V_{\rm 1}$, $V_{\rm 2}$) across a resistance with a standard lock-in technique.
This structure works as a \textit{pure} two-path interferometer when the tunnel-coupled wires are properly tuned to half beam splitters \cite{Yamamoto:2012fk, Tobias2014, Aharony2014, TakadaAPL2015}, where the two output currents oscillate in anti-phase as plotted in Fig.\,\ref{fig:device}(b).
All measurements were performed in a dilution refrigerator at the base temperature of about $70$ mK except for Fig.\,\ref{fig:Kondo}(a).

%%%% Characterization of the Kondo correlation %%%%%%%%%%%
%
%%%%%%%%% Figure 2 %%%%%%%%%%%%%%%%%%%%
%%%%%%%%%%%%%%%%%%%%%%%%%%%%%%%%%%%
We start by characterizing the Kondo correlation of the QD by completely depleting the regions beneath both tunnel-coupling gates. 
As a consequence all the injected current passes through the QD and is recovered at the lower left contact ($I_{\rm 1}^*$ and $G_{\rm 1}^*$).
Figure\,\ref{fig:Kondo}(a) shows the temperature dependence of the CPs.
Kondo correlations appear at the valley around $V_{\rm p} = -0.56$ V, where the linear conductance logarithmically increases as the temperature is decreased below $200$ mK.
Since we do not reach the unitary limit, the Kondo temperature cannot be determined precisely.
Instead we can estimate from the temperature dependence of the conductance a lower bound of $T_{\rm K}$ of $100$ mK at the valley center, below which the conductance exceeds $e^2/h$.
In addition, we measure a Coulomb diamond for this parameter set and confirm the so-called zero bias anomaly, a typical feature of Kondo physics [Fig.\,\ref{fig:Kondo}(b)].
The horizontal yellow ridge in Fig.\,\ref{fig:Kondo}(b) is an indication of the resonant DOS near the Fermi level.

%%%% Phase measurement at T << T_K %%%%%%%%%%%%%%%
%%%%%%% Figure 3 %%%%%%%%%%%%%%%%%%%%
\begin{figure}[htbp]
	\includegraphics[width=0.48\textwidth]{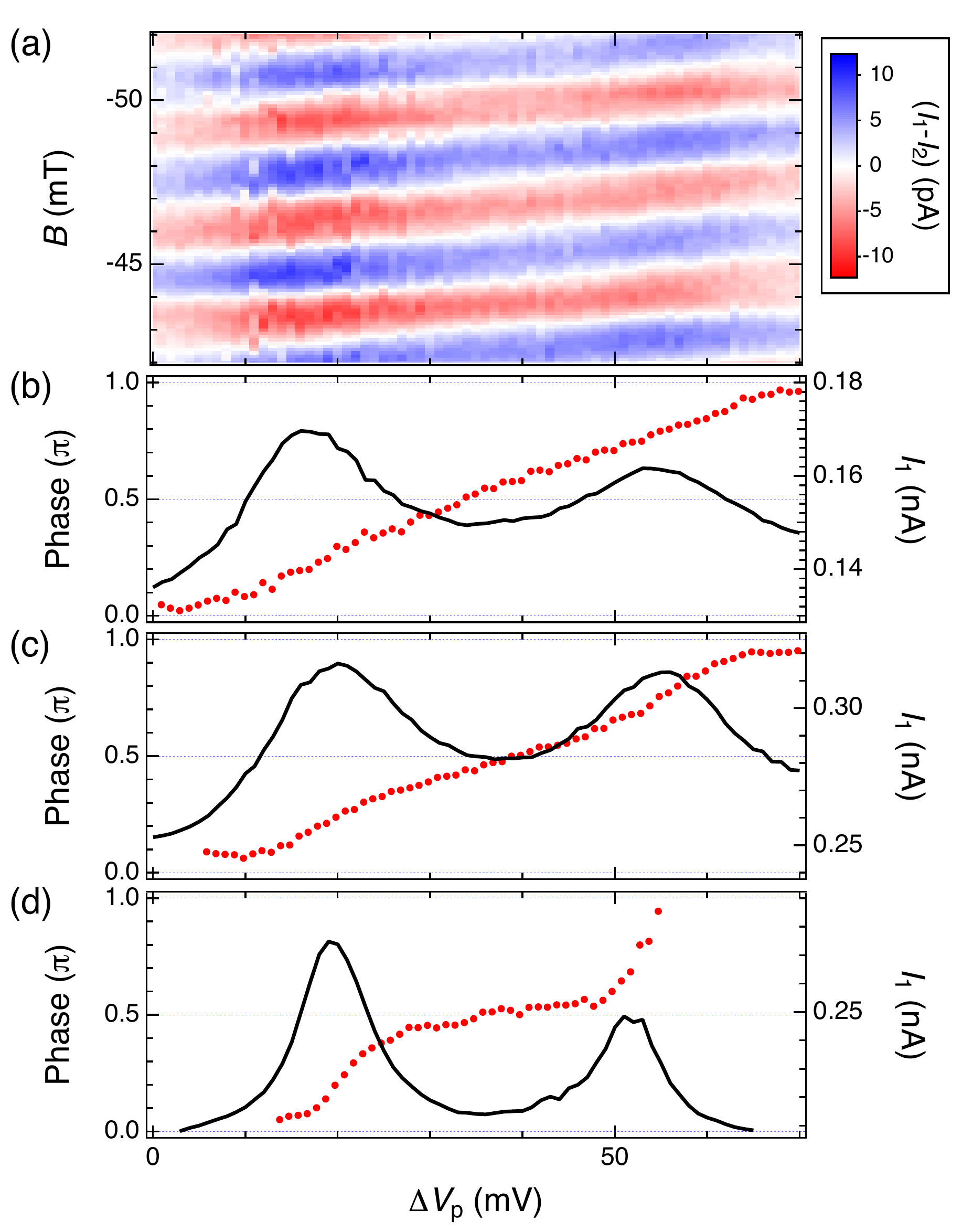}
	\caption{\label{fig:SmoothShift}(a) Magneto oscillations of ($I_{\rm 1} - I_{\rm 2}$) as a function of $V_{\rm p}$ across the two Coulomb peaks showing Kondo correlations. The non-oscillating background as a function of the magnetic field is subtracted. (b)-(d) The transmission phase shift (red circle, left axis), determined by a complex FFT of the magneto oscillations, together with $I_{\rm 1}$ (black line, right axis). $I_{\rm 1}$ is averaged over one oscillation period of the magnetic field. From (b) to (d) the coupling between the QD and the leads is reduced.}
\end{figure}
%%%%%%%%%%%%%%%%%%%%%%%%%%%%%%%%%%
In order to measure the transmission phase shift across these CPs we retune both tunnel-coupled wires to half beam splitters.
We record the magnetic field dependence of the two output currents at different values of $V_{\rm p}$ when scanning across the peaks.
Figure~\ref{fig:SmoothShift}(a) shows the result, where the difference of the currents $I_{\rm 1}$ and $I_{\rm 2}$ oscillating with opposite phase is plotted after subtraction of the non-oscillating smoothed background \cite{Takada2014}.
The phase of the magneto-oscillation smoothly shifts as a function of $V_{\rm p}$ across the peaks with Kondo correlations.
In Fig.~\ref{fig:SmoothShift}(b) we plot the phase shift obtained by a complex fast Fourier transform (FFT) of the data of Fig.~\ref{fig:SmoothShift}(a), together with the current $I_{\rm 1}$ averaged over one oscillation period for the magnetic field.
The current mimics the shape of the CPs on top of the background current coming from the right path of the interferometer.
The phase smoothly shifts by $\pi$ across the two CPs without showing any plateau feature, as already visible in Fig.~\ref{fig:SmoothShift}(a).

Next, we slightly reduce the coupling $\Gamma$ between the QD and the leads  by suitably tuning the gate voltages of the QD and measure again the phase shift [Fig.~\ref{fig:SmoothShift}(c)].
One observes a slight change in the slope of the phase shift, in particular, in the valley between the two CPs.
By further reducing $\Gamma$ the phase is clearly locked around $\pi/2$, as shown in Fig.~\ref{fig:SmoothShift}(d).
For all these gate conditions shown in Fig.~\ref{fig:SmoothShift} the temperature $T$ should still be lower than $T^{\rm c}_{\rm K}$, the Kondo temperature at the center of the Kondo valley, since the phase shift across one of the two CPs does not exceed $\pi/2$.
If one reduces $\Gamma$ even further so that $T^{\rm c}_{\rm K}$ drops below $T$, the phase shift  should exceed $\pi/2$ at the left of the Kondo valley and should decrease below $\pi/2$ to right of the Kondo valley, resulting in an $S$-shaped phase evolution as recently demonstrated  in Ref. \onlinecite{Takada2014}.
Although such a crossover from $T^{\rm c}_{\rm K} > T$ to $T^{\rm c}_{\rm K} < T$ at the center of the Kondo valley could not be observed for these two specific CPs due to limitations in the tuning parameters, we confirmed such a behavior for different peaks with Kondo correlations in the current device.

\begin{figure}[htbp]
	\includegraphics[width=0.45\textwidth]{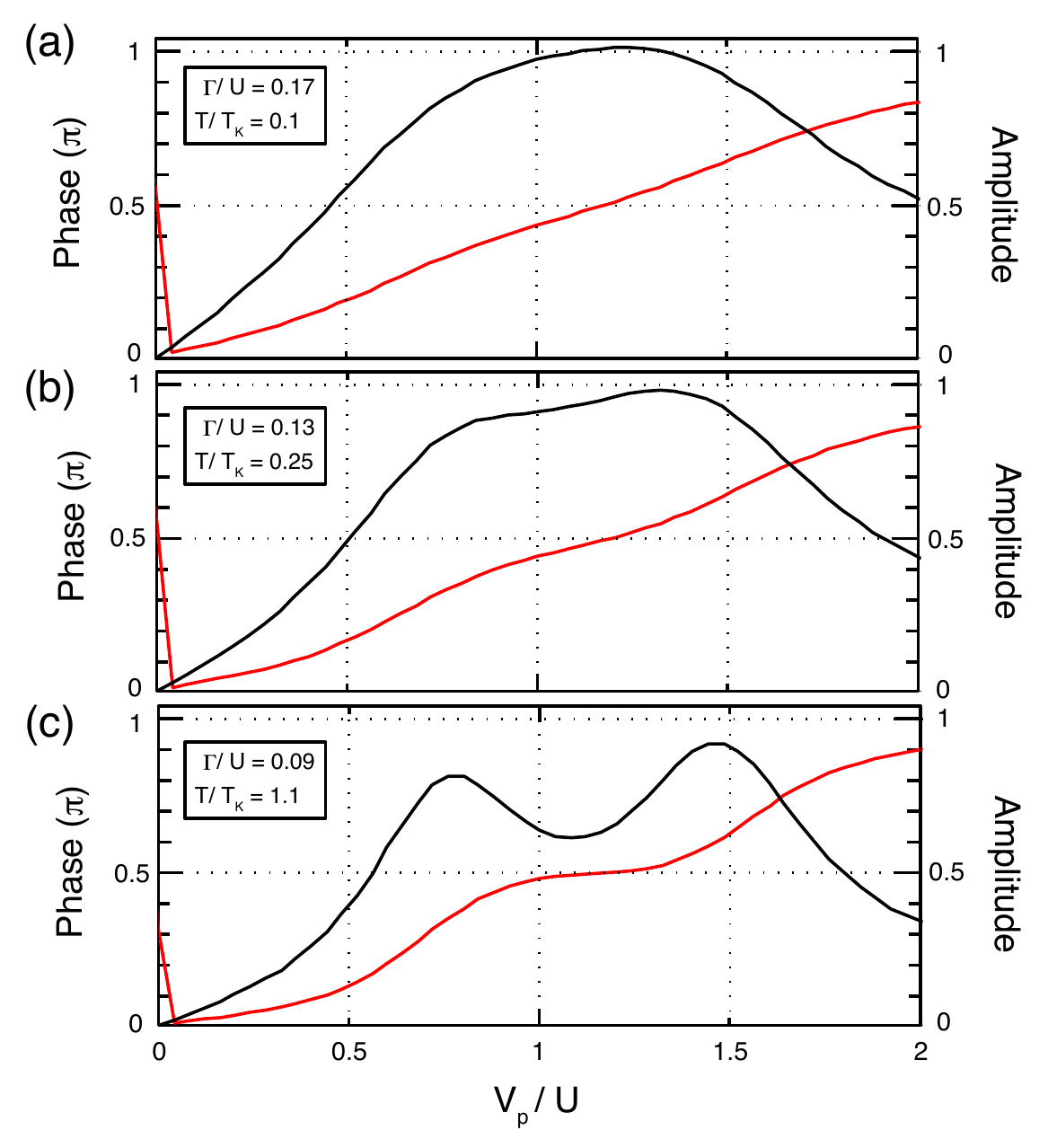}
	\caption{\label{fig:NRG}Transmission phase (red line, left axis) and transmission amplitude (black line, right axis) calculated by a two-level Anderson impurity model. The single-level spacing is fixed to 0.3~U and the orbital parity relation between two single-particle levels is chosen to be same. The phase shift across two CPs, which originates from the upper energy level is shown here. The parameters used for the calculations are indicated in the graphs. $T^{\rm c}_{\rm K}$ indicates the Kondo temperature at the center of the Kondo valley.}
\end{figure}

To compare such a behavior with theory, we show NRG calculations employing a two-level Anderson impurity model \cite{Wilson1975, Weichselbaum2007, Hecht2009, Weichselbaum2012b} using the same model assumptions as in Ref.\,\onlinecite{Takada2014}.
Following the experiment, $\Gamma$ is changed as a parameter while temperature, single-level spacing, and the orbital parity relation between the two single-particle levels are fixed.
For simplicity, $\Gamma$ is set to be equal for the two levels and the $V_{\rm p}$-dependence of the parameters is not taken into account.
The energy scale of the parameters is characterized in units of charging energy $U$ .
We indeed find a similar phase behavior in the calculations, as shown in Fig.\,\ref{fig:NRG}.
For the largest ${\rm \Gamma/U}$ [Fig.\,\ref{fig:NRG}(a)] the phase evolves almost linearly by $\pi$.
When ${\rm \Gamma/U}$ is reduced, the slope of the phase starts to decrease near the center of the valley [Fig.\,\ref{fig:NRG}(b)].
By further reducing ${\rm \Gamma/U}$ a phase plateau at $\pi/2$ develops \cb{[Fig.\,\ref{fig:NRG}(c)]}.
Even though the parameters of the calculations are not quantitatively fit to the experimental conditions, which results in different transmission amplitudes from those of the experiment, the NRG calculations reproduce the experimentally observed phase evolution.

The smooth phase shift without the $\pi/2$ plateau observed in Figs.\,\ref{fig:SmoothShift}(a)-\ref{fig:SmoothShift}(c) and reproduced in Figs.\,\ref{fig:NRG}(a) and \ref{fig:NRG}(b) is quite remarkable.
Usually one associates the Fermi-liquid ground state with a phase shift plateau at $\pi/2$,  a hallmark of the Kondo singlet ground state.
In the above experiment and calculation, however, the plateau at $\pi/2$ only develops when $\Gamma/U$ is reduced [Figs.\,\ref{fig:SmoothShift}(d) and \ref{fig:NRG}(c)].
%%%%%%%% Figure 4 %%%%%%%%%%%%%%%%%%%%
%%%%%%%%%%%%%%%%%%%%%%%%%%%%%%%%
Such a behavior can be understood in terms of the Friedel sum rule and thermal averaging of the phase.
To see this, we evoke a recently-developed Fermi-liquid theory of Ref.\,\onlinecite{Mora2015}, which generalizes the Fermi-liquid theory that Nozi\`{e}res had developed many years ago for the Kondo model to the case of the Anderson impurity model, including the case of particle-hole asymmetry.
According to Ref.\,\onlinecite{Mora2015}, the transmission phase shift at zero temperature and magnetic field depends \textit{linearly} on energy for sufficiently small excitation energies $\varepsilon$ with respect to the Fermi energy.
\begin{equation}
	\delta (\varepsilon) = n_d \cdot \pi/2 + \varepsilon/E^*\ \ {\rm for}\ |\varepsilon| \ll E^*
	\label{eq:1}
\end{equation}
The first term reflects the Friedel sum rule, with $n_d$ being the ground state expectation value of the dot occupancy.
The second term parametrizes the leading linear energy dependence of the phase in terms of a characteristic low-energy scale $E^*$.
Both $n_d$ and $E^*$  depend on gate voltage, with $E^*$ minimal at the center of the Kondo valley.
If the system is in the \textit{local moment regime}, defined by the requirements that $n_d \simeq 1$ and that charge fluctuations are negligible, then $E^*$ is proportional to the Kondo temperature $T_{\rm K}$.
Outside the local moment regime, where $n_d$ differs significantly from 1 and/or charge fluctuations are strong, the phenomenology typically associated with the Kondo effect, such as the existence of a well-defined Kondo resonance, is no longer applicable.
Nevertheless, Eq.\,(\ref{eq:1}) still holds, although $E^*$ no longer has the interpretation of a Kondo temperature.
For large excitation energies, $|\varepsilon| \gg E^*$, the phase $\delta (\varepsilon)$ depends nonlinearly on energy [as shown, e.g., in Figs.\,2(b) and 2(d) of Ref.\,\onlinecite{Gerland2000}].

For sufficiently small temperatures ($T \ll E^*/k_{\rm B}$), the transmission phase $\delta_{\rm expt}$ measured in our experiment can heuristically be associated with a thermal average of $\delta(\varepsilon)$ over an energy window of width $k_{\rm B}T$ [see Eq.~(A22) of Ref.\,\onlinecite{Hecht2009}].  Since the contribution of the linear term vanishes upon thermal averaging, we obtain
\begin{equation}
	\delta_{\rm exp} \simeq n_d \cdot \pi/2
	\label{eq:2}
\end{equation}
thus the measured phase can directly be interpreted in terms of the Friedel sum rule.
We interpret the linear change of $\delta_{\rm expt}$ with gate voltage observed in Figs.\,\ref{fig:SmoothShift}(b) and
\ref{fig:SmoothShift}(c) as evidence that the dot occupancy changes essentially linearly with gate voltage.
This would imply that although the conductance does show some traces of Kondo correlations (Fig.\,\ref{fig:Kondo}), the ratio of $\Gamma/U$ is so large that a well-defined local moment regime, with $n_d \simeq 1$ over a range of gate voltages, is not realized.

If $\Gamma/U \ll 1$, then a well-defined local moment with $n_d \simeq 1$ is realized throughout the Kondo-valley range of gate voltages ($-U + \Gamma \lesssim \varepsilon_d \lesssim -\Gamma$).
For $T \ll T_K^{\rm c}$ (which implies $T \ll E^*/k_{\rm B}$ throughout the Kondo valley), Eq.\,\eqref{eq:2} leads to the well-known phase plateau $\delta_{\rm expt} \simeq \pi/2$.
Figure\,\ref{fig:SmoothShift}(d) is close to this situation.
However, it is important to recall that the applicability of Eq.\,\eqref{eq:2} is limited solely to $T \ll E^*/k_{\rm B}$.
Since smaller $\Gamma/U$ leads to smaller $E^*$ for the fixed gate voltage $V_{\rm p}$, in Fig.\,\ref{fig:SmoothShift}(d) the temperature is not small compared to $E^*$, but rather comparable to it ($T \simeq E^*/k_{\rm B}$) around the center of the valley.
Therefore, for Fig.\,\ref{fig:SmoothShift}(d) the measured phase $\delta_{\rm expt}$ is not given simply by the Friedel sum rule
  \eqref{eq:2}, since for such comparatively large temperatures, contributions to $\delta(\varepsilon)$ that are nonlinear in energy [not shown in \eqref{eq:1}] begin to contribute.
If $T \simeq E^*/k_{\rm B}$, this results in a $\pi/2$-phase plateau at the center of the Kondo valley that is more pronounced than the plateau expected from Eq.\,(\ref{eq:2}).
We have confirmed this by NRG calculations in which either $T/U$ or $\Gamma/U$ are varied: Irrespective of the detailed values of the parameters, a well-defined $\pi/2$-phase plateau is found at temperatures of order or even somewhat above $T_K^{\rm c}$.
This is consistent with previous findings in the literature, e.g., Fig.\,4(b) of Ref.\,\onlinecite{Gerland2000}, Fig.\,1 of
Ref.\,\onlinecite{Silvestrov2003}, and Figs.\,1(a) and 1(b) of Ref.\,\onlinecite{Hecht2009}.
This is also consistent with Fig.\,\ref{fig:SmoothShift}(d), where the phase is locked around $\pi/2$ when the QD is far from the unitary limit.

The enhancement of the $\pi/2$-phase plateau can be understood by the following heuristic argument.  For this, we need to consider the resonance state formed near the Fermi energy, which is the dominant contribution for the transport at $T < E^*/k_{\rm B}$.
As discussed above, the experimentally observed phase shift is a thermal average over the energy window of $k_{\rm B}T$ and it is this thermal averaging with its respective weight (transmission probability) of the resonance DOS.
The center of the resonance state is misaligned from the Fermi energy except at the center of the Kondo valley due to the
strong charge fluctuation, as seen from the almost linear phase evolution for $T \ll E^*/k_{\rm B}$.
Resonant tunneling leads to a $\pi/2$-phase shift, while tunneling with energy above (below) the resonance leads to a larger (smaller) phase shift compared to $\pi/2$.
When the resonance enters the $k_{\rm B}T$ energy window, the weight of the tunneling probability is maximal at the resonance, while it decreases for energies above and below resonance.
As a consequence, the $\pi/2$-phase shift should be enhanced by the thermal average.
Even though nonlinear energy dependent phase contributions exist at $T \simeq E^*/k_{\rm B}$, this picture should be valid as far as the electron transport is dominated by the resonance DOS.
For larger $T$, contributions from the single-particle DOS cannot be ignored, which will cause $\delta_{\rm expt}$ to show an $S$-shaped behavior as a function of gate voltage, as predicted in Fig.\,3(b) of Ref.\,\onlinecite{Gerland2000} and observed experimentally in Ref.\,\onlinecite{Takada2014}.
Such a crossover of the transport regime should occur at $T \simeq E^*/k_{\rm B}$ at the center of the Kondo valley.
Hence we can estimate $E^*/k_{\rm B}$ (which corresponds to $T_{\rm K}$ in the local moment regime,) from the
observation of the transition of the phase behavior.

In summary, we studied the phase behavior of a Kondo correlated QD formed in a GaAs 2DEG for $T \le T^{\rm c}_{\rm K}$.
We observed that the phase smoothly shifts by $\pi$ across two consecutive CPs without showing any plateau for large $\Gamma$.
Only when $\Gamma$ is decreased does the phase lock at $\pi/2$ near the center of the Kondo valley and develops a plateau.
Such a behavior is reproduced by NRG calculations and can be understood in terms of the Friedel sum rule and a heuristic argument of thermal average.
The $\pi/2$-phase shift is pronounced at $T \simeq E^*/k_{\rm B}\ (\simeq \ T_{\rm K})$ due to thermal average, which allows one to estimate the transition temperature.
Our results demonstrate that the phase shift in a Kondo correlated QD is much more subtle than naively expected for a localized Kondo impurity.

S. Takada acknowledges financial support from the European Unions Horizon 2020 research and innovation program under the Marie Sklodowska-Curie grant agreement No 654603.
M.Y. acknowledges the financial support by Grant-in-Aid for Young Scientists A (No. 23684019), Grant-in-Aid for Challenging Exploratory Research (No. 25610070) and a Grant-in-Aid for Scientific Research A (No. 26247050).
C.B. acknowledges financial support from the French National Agency (ANR) in the frame of its program BLANC FLYELEC project No. anr-12BS10-001, as well as from DRECI-CNRS/JSPS (PRC 0677) international collaboration.
A.A. and J. v. D. acknowledge support from the DFG through SFB-631, SFB-TR12, and the Cluster of Excellence Nanosystems Initiative Munich.
A.L. and A.D.W. acknowledge gratefully the support of Mercur Pr-2013-0001, DFG-TRR160, BMBF-Q.com-H 16KIS0109, and the DFH/UFA CDFA-05-06.
S. Tarucha acknowledges the financial support by JSPS, Grant-in-Aid for Scientific Research S (No. 26220710), MEXT project for Developing Innovation Systems, and QPEC, the University of Tokyo.
%
%%%%%%%%%%%%%%%%%%%

\end{document}